\documentstyle[aps]{revtex}

\input{tcilatex}

\begin{document}
\title{Vortex dynamics differences in YBaCuO at low temperatures for $H||ab$ planes
due to twin boundary pinning anysotropy.}
\author{S. Salem-Sugui, Jr.$^{1}$, A. D. Alvarenga$^{2}$, M. Friesen$^{3}$, K. C.
Goretta$^{4}$, O. F. Schilling$^{5}$, F. G. Gandra$^{6}$, B. W. Veal$^{7}$,
P. Paulikas$^{7}$}
\address{$^{1}$Instituto de F\'{\i }sica, Universidade Federal do Rio de Janeiro\\
C.P.68528, 21945-970 Rio de Janeiro, RJ, Brasil\\
$^{2}$Centro Brasileiro de Pesquisas Fisicas, Rua Dr. Xavier Sigaud,150,\\
22290-180 Rio de Janeiro, RJ, Brazil\\
$^{3}$Department of Materials Science and Engineering, 1500 Engineering\\
Drive, University of Wisconsin, Madison, Wisconsin 53706\\
$^{4}$Energy Technology Laboratory, Argonne National laboratory, Argonne,\\
Illinois\\
60439-4838\\
$^{5}$Departamento de F\'{i}sica, Universidade Federal de Santa Catarina,\\
88040-900 Florian\'{o}polis,SC, Brazil.\\
$^{6}$Instituto de Fisica, UNICAMP, CP6185, 13083-970 Campinas, SP, Brazil\\
$^{7}$Materials Sciences Division, Argonne National Laboratory, Argonne,\\
Illinois, 60439-4838}
\maketitle

\begin{abstract}
We measured magnetization, M, of a twin-aligned single crystal of YBa$_{2}$Cu%
$_{3}$O$_{x}$ (YBaCuO), with Tc = 91 K, as a function of temperature, T, and
magnetic field, H, with H applied along the ab planes. Isothermal M-vs.-H
and M-vs.-time curves were obtained with H applied parallel ($\parallel $)
and perpendicular ($\perp $) to the twin boundary, TB, direction. M-vs.-H
curves exhibited two minimums below 38 K, which resembled similar curves
that have been obtained in YBaCuO for H$\parallel $c axis. Above 12 K, the
field positions of the minimums for H$\parallel $TB and H$\perp $TB were
quite similar. Below 12 K, the position of the second minimum, Hmin,
occurred at a higher field value with H$\parallel $TB. Below 6 K, only one
minimum appeared for both field directions. At low temperatures, these
minimums in the M-vs.-H curves produced maximums in the critical current. It
was determined that vortex lines were expelled more easily for H$\parallel $%
TB than for H$\perp $TB, and, therefore, below a certain field value, that J$%
_{c}$(H$\perp $TB) was larger than J$_{c}$(H$\parallel $TB) . At T 
\mbox{$<$}%
12 K with H$\parallel $TB, the relaxation rate for flux lines leaving the
crystal was found to be different from that for flux entering the crystal.
We also observed flux jumps at low temperatures, with their sizes depending
on the orientation of magnetic field with respect to the TBs.
\end{abstract}

$Introduction$

Twin planes are ubiquitous in the high-temperature superconductor YBa$_{2}$Cu%
$_{3}$O$_{x}$(YBaCuO). Under microscopic analysis they appear as flat,
slab-like domains of micrometer thickness. The domains are coherent and are
oriented in various $[110]$ directions. The a and b axes are inverted in
neighboring domains. The domain boundaries are commonly referred to as twin
boundaries (TBs). The boundary region has a structure different from that of
the bulk crystal, including a relative deficiency of oxygen \cite
{stern,bakel,liu} or excess of impurities that can accumulate during crystal
growth. TBs form strong vortex pinning centers and are responsible for a
rich variety of transport anisotropies. Under typical growth conditions,
these anisotropies cannot be observed readily because the neighboring TBs
are not aligned. The gross vortex dynamics of samples with such TBs are
determined by the sample's texture. Twin-aligned YBaCuO samples can,
however, be formed, and such samples provide windows into the effects on
superconducting properties of TBs.

Initial studies of twin-aligned YBaCuO single crystals revealed strong
anisotropies for magnetic fields perpendicular to the c axis of the crystal,
H$\perp $c, for TBs oriented either parallel or perpendicular to the field 
\cite{liu,swartzendruber,kwok,duran}. These studies included isothermal
measurements of resistivity and magnetic hysteresis. Additional angular
dependencies between H and TBs were reported in [\onlinecite{gyorgy}] and [%
\onlinecite{ousseana}]. The effects of TBs on vortex dynamics were studied
by magneto-optical measurements for H parallel and perpendicular to the c
axis \cite{vlasko-vlasov,rinke,duran2}. More recent studies of twin-aligned
single crystals of YBaCuO include transport measurements as a function of
current in the ab plane \cite{pastoriza} and Bitter decoration experiments
under tilted magnetic fields \cite{herbsommer}. Vortex-pinning \cite{jorge}
and flux-creep measurements have been studied through use of ac probes for
the configurations H parallel to the TBs, H$\parallel $TB, and H
perpendicular to the TBs, H$\perp $TB \cite{bondareko}.

Previous TB-based studies of YBaCuO focused mainly on higher temperatures,
which motivated the present work in the temperature region 50 K 
\mbox{$<$}%
T 
\mbox{$<$}%
2 K. We studied the effects of TB pinning anisotropy on magnetic-hysteresis
and magnetic-relaxation curves for a twin-aligned single crystal of YBaCuO,
with the magnetic field applied in the ab plane for two directions, namely H$%
||$TB and H$\perp $TB. The study revealed interesting features in magnetic
hysteresis and magnetic relaxation curves due to TB vortex-pinning
anisotropy, that, to our knowledge, have not yet been reported: We observe
the existence of a second minimum in magnetic hysteresis curves below 38 K,
which temperature behavior below 12 K for $H||TB$ is quite different than
the one observed when $H\perp TB$. The temperature behavior (below 12 K) of
the relaxation rate for $H||TB$ (studied for H = 3 T), is also quite
distinct of the behavior observed when $H\perp TB$. At low temperatures,
maximums in the critical current J$_{c}$ occurred at the positions of the
minimums in the $M(H)$ curves. We also observe flux-jumps in the
magnetization curves at low temperatures, with sizes depending wether H is
applied parallel or perpendicular to TB.

Flux jumps result from thermomagnetic instabilities associated with
dissipative heating (either flux flow or avalanche). If the dissipative
heating cannot diffuse through the sample, it can increase the local
temperature, possibly even above the critical temperature, T$_{c}$,
producing a jump in magnetization. Such jumps may occur in response to
changes in the external field if the magnetic-diffusion time is shorter than
the thermal-diffusion time. The size of a jump depends on the rate of
magnetic field increase, dH/dt. Theoretically, the stability criterion \cite
{esquinazi,mints} defines a critical thickness below which flux jumps do not
occur. Experimentally, such results have been confirmed for melt-textured
YBaCuO with the magnetic field applied perpendicular to the c axis, H$\perp $%
c \cite{muller}. Additional asymmetries have been observed in the different
branches of magnetization hysteresis curves, M(H), for melt-textured YBaCuO
at T 
\mbox{$<$}%
6 K, with a greater prevalence of flux jumps occurring with increasing field
than with decreasing field \cite{muller}. These results are in contrast to
those for H$||$c, for which no hysteresis asymmetries have been observed 
\cite{muller,guillot}. For practical applications, especially those
involving high currents, characterization and management of flux jumps
become critical. We therefore analyzed the observed flux jumps in some
detail.

$Experimental$ $details$

The sample was a single crystal of YBaCuO with T$_{c}$ = 91 K and dimensions 
$\sim $1 x 1 x 0.1 mm. Figure 1 shows an enlarged (100x) photo of the
sample's surface. All twin boundaries within the crystal were parallel and
extended across the entire thickness of the sample, as confirmed by
microscopic analysis. The density of TBs was estimated to be 45 twins per
mm. The sample had an approximately square shape, with one slightly rounded
corner, and the twin boundaries displaced perpendicular to the larger
diagonal.

Magnetization and magnetic-relaxation data were taken after cooling the
sample in zero applied magnetic field. A commercial magnetometer (Quantum
Design PPMS-9T) was utilized for the measurements. The magnetic signal of
the sample (plus sample holder) was obtained from the inductive signal of a
pick-up coil, which appeared because of motion of the sample through the
coil in a homogeneous magnetic field, which we term a scan. Each set of
magnetization data represents the average of three scans.
Magnetization-vs.-field, $M(H)$, curves were obtained at fixed temperatures
ranging from 2 to 50 K. For a fixed applied field of H = 3 T,
magnetic-relaxation measurements were obtained at 60 s intervals over a
period of 3600 s, for both the upper and lower branches of a hysteresis
curve. The remanent magnetization at zero field was also recorded. The value
of 3 T was chosen to minimize the effects of field penetration during the
measurements \cite{reply}. We refer to the increasing field magnetization as 
$Min$, the decreasing field magnetization as $Mout$, and the remanent
magnetization as $Mrem$. The latter two signals were obtained after first
increasing the field to 9 T.

In the temperature range 2-12 K, all relaxation measurements were obtained
for both field orientations with respect to the TBs. Care was taken to
assure that the magnetic field was applied in the ab plane. The sample was
mounted on a flat surface machined into the center of a 3 cm wooden cylinder
that fit snugly into a straw that was inserted into the magnetometer. An
optical microscope with polarized light was used for sample mounting. The
angle between the TBs and the magnetic field H was estimated to be accurate
to 
\mbox{$<$}%
$2{{}^{\circ }}$. After the experiment was concluded, we measured $M(H)$ for
the sample holder at all relevant experimental temperatures to account for
background corrections.

$Results$ $and$ $discussion$

Figure 2 contains selected M-vs.-H curves obtained for both orientations of
H with respect to the TBs at T = 25, 15, and 8 K. The arrows in Fig. 2 are
pointed at the curves obtained at 8 K. From left to right in the figure,
they are: the first arrow is the first minimum $Hpen$, which is associated
to field penetration; the second arrow is the local maximum $Hon$, which is
associated with the field at which pinning sets in; and the third and forth
arrows are the second minimums $Hmin$, which are reminiscent of the second
magnetization peaks observed in high-T$_{c}$ superconductors for H$||$c
axis. The decreasing-field portions of M-vs.-H curves do not show any
maximum. The third and forth arrows show, respectively, Hmin for H$\perp $TB
and Hmin for H$||$TB. These three fields--$Hpen$, $Hon$, and $Hmin$--were
clearly present in all M-vs.-H curves obtained at 8-33 K.

M-vs.-H curves obtained at or above 38 K exhibited only a single minimum,
which is associated with $Hpen$. M-vs.-H curves obtained below 8 K also
exhibited only one minimum, which, in this particular case, may possibly be
associated with $Hmin$. This conjecture will be discussed below.

The positions of $Hpen$, $Hon$, and $Hmin$ for both directions of applied
field were approximately the same for temperatures 12 K $<$ T $<$ 38 K, but
a change in the field position of $Hmin$ was noted below 12 K. Such a change
in field position can be observed in the $M(H)$ curves at 8 K depicted in
Fig.2, in which $Hmin$ for H$||$TB occurred at a much higher field than did $%
Hmin$ for H$\perp $TB. The physical reason for the shifting of $Hmin$, as
shown at 8 K, to occur only below 12 K is not clear. It may be related to
the temperature behavior of the TB barriers found below 12 K, as will be
discussed below. The insets in Fig.2 show the angular dependence at 8 K of $%
M(H)$ for small angles. These data will also be discussed below.

Figure 3 shows $M(H)$ curves obtained at 2 K in the main figure and at 4 K
in the inset. These curves clearly show the existence of a single minimum.
As observed for $Hmin$ at 8 K, the field position of the minimum is higher
for H$||$TB than for H$\perp $TB, which suggests that the minimum observed
below 8 K is associated with $Hmin$ rather than with $Hpen$. The inflection
points apparent in the curves of Fig. 3 for fields $H$ $<$ $Hmin$, which are
most visible in the H$\perp $TB curves, are possibly related to $Hon$. Flux
jumps are also evident for H$||$TB. A smaller flux jump was also observed at
4 K (inset of Fig. 3) for H$||$TB only.

The principal differences, that are due to TBs, among the curves in Fig. 2
and between those in Fig 3 at fixed temperatures are as follows.

(1) The values of $\Delta M$ for intermediate and higher fields were higher
for H$||$TB than for H$\perp $TB. This result been obtained before; it is
due to twin-boundary pinning \cite{liu}.

(2) The diamagnetic signal for $H$ $<$ $Hpen$ was higher for H$||$TB than
for H$\perp $TB. This result has been also observed previously \cite{liu}.
For the same applied field value H, differences in the diamagnetic signal in
the field-penetration region, just above H$_{c1}$, suggest that the local
field $H_{i}$ at the sample surface is smaller when H$\perp $TB than when H$%
||$TB. Such a result is possible if the demagnetization factor of the sample
is higher for the configuration H$\perp $TB: $H_{i}$ = H- NM, where N is the
demagnetization factor. Our sample was a thin slab and the experiment was
conducted with the magnetic field lying in the plane of the slab. This
configuration suggests that demagnetization fields were quite small. On the
other hand, although the crystal's face was approximately square, one corner
was rounded, with the twin boundaries displaced perpendicular to the larger
diagonal. Therefore, the sample geometry for H$||$TB was significantly
different than for H$\perp $TB, and the demagnetization factor for each case
would be expected to be different, which may explain the differences in the
diamagnetic signal observed for each case.

(3) After decreasing the field until H = 0 (the decreasing-field branches of
the M-vs.-H curves), the remanent magnetization defined as M(H = 0) was
higher for H$\perp $TB than for H$||$TB \cite{liu}. In fact, magnetization
in the decreasing-field branch started to become higher for H$\perp $TB
below a certain field, the value of which increased as temperature
decreased. This response was observed for all $M(H)$ curves and can be
clearly observed in Fig. 4, in which the estimated critical current density, 
$J_{c}$, vs. field is plotted for temperatures below 10 K, for both field
directions with respect to the TBs. $J_{c}$ in A/cm2 was estimated from the
Bean critical-state model \cite{beam}. Below a certain field, $J_{c}$ for H$%
\perp $TB was always higher than $J_{c}$ for H$||$TB; this response was
observed for all curves from 50 K to 2 K. This relationship between $J_{c}$
and the TBs does not appear to have been discussed in the literature. It is
interesting to note that as temperature was lowered, at a field of a few
Tesla a broad maximum in $J_{c}$ emerged (maximums were clear at 4 K). The
position of each maximum in $J_{c}$ appeared to be related to the respective
field position of the minimum ($Hmin$) in each $M(H)$ curve at 4 K.

After obtaining the data set at 2-50 K, we measured $M(H)$ curves at 8 K,
with the applied magnetic field tilted within the plane by a small angle (5${%
{}^{\circ }}$ 
\mbox{$<$}%
$\theta $ 
\mbox{$<$}%
10${{}^{\circ }}$) relative to the original directions, H$\perp $TB ($\theta 
$ = 90${{}^{\circ }}$) and H$||$TB ($\theta $ = 0${{}^{\circ }}$). (The
magnetic field remained in the ab plane when the sample was rotated.)
Although a full set of angular-dependence measurements was beyond the scope
of this work, this limited set of measurements allowed us to check for
possible edge (or geometric) effects in the $M(H)$ curves at temperatures at
which two minimums had been clearly resolved. The results for each set of $%
M(H)$ curves are shown in the insets of Fig. 2. Inspection of the insets
reveals two facts:

(a) Rotation of the TBs by a small angle with respect to the applied field
changed the position of the second minimum for H$||$TB ($Hmin$ was displaced
by a small field value after the sample was rotated), but not for H$\perp $%
TB. Change of the position of $Hmin$ only for the case H$||$TB suggests that 
$Hmin$ is related to TB pinning anisotropy. We further speculate that the
absence of the corresponding second peak in the decreasing-field branch of
the curve may be due to the fact that with decreasing of the field, flux
lines could leave the sample relatively easily through the TBs.

(b) Rotation of TBs with respect to the applied field by a small angle had
considerable effect on the shape of the $M(H)$ curves in the
field-penetration region ($Hpen$ $<$H$<$ $Hon$) for both H$\perp $TB and H$%
|| $TB, which provides evidence for the importance of edge (or geometric)
effects in this field region. The $M(H)$ curve for $\theta $ = 7${{}^{\circ }%
}$ (left inset of Fig. 2) is apparently rotated with respect to the $M(H)$
curve for $\theta $ = 0${{}^{\circ }}$ in the same figure. This apparent
rotation in the $M(H)$ curve is simply due to a change in the position of
the sample in the holder: After rotating the sample, it was displaced out of
the middle of the its holder, and the signal due to the sample holder (i.e.,
the background magnetization) was not subtracted correctly, which produced
the apparent rotation. In our experimental set-up, the sample was fixed to
the sample holder with GE varnish. Changing the sample's position
necessitated diluting the varnish, rotating the sample, and then reattaching
the sample at the exact correct position. A full angular-dependence
experiment would require the sample to be fixed to an appropriate rotator,
to avoid the possibility of sample damage.

Differences between the curves with H$||$TB and H$\perp $TB of Figs. 2 and
3, as listed above in itens 1 and 3, can be explained by the following
considerations. (1) There is a vortex line along the magnetic field
direction; i.e, there are no pancake-vortices in the $ab$ planes. (2)
Defects located at the boundaries of the twin planes act as pinning centers
and also prevent vortices from crossing the TBs when H$||$TB. (3) For
intermediate and higher fields, the density of vortex lines pinned with
increasing and decreasing applied field is higher for H$||$TB than for H$%
\perp $TB. This final contention can be obtained after comparison of the
values of magnetization in the increasing (decreasing) field of a curve
obtained for H$||$TB with the correspondent (at same temperature) curve
obtained for H$\perp $TB, as well by comparing the respectives values of $%
\Delta M(H)$ of both curves. As a consequence of the preceding facts, the
average distance between vortex lines, for intermediate and higher fields,
is smaller for H$||$TB than for H$\perp $TB.

The vortex-vortex interaction energy \cite{Tinkham} is given by $%
F_{12}=(\phi _{0}^{2}/8\pi ^{2}\lambda ^{2})K_{0}(r_{12}/\lambda )$, where $%
\phi _{0}$ is the quantum flux, $\lambda $ is the penetration depth, $r_{12}$
is the average distance between vortex lines, and $K_{0}$ is a zeroth-order
Hankel function of imaginary argument. The interaction given by $F_{12}$ is
repulsive, and the repulsive force between vortex lines (given by $-\partial 
$ $F_{12}/\partial x$ for the $x$ direction) increases as $r_{12}$
decreases. Imbalance between $-\partial $ $F_{12}/\partial x$ and the
magnetic pressure may produce vortex motion which, in the case of decreasing
of the field, produces vortex exit. From the above considerations, the
repulsive force is higher for H$||$TB than for H$\perp $TB.

By assuming a triangular lattice of vortex lines  (it is also assumed that
only the boundary regions of the TBs plane can stronly pin a vortex) ,
the repulsion between vortex lines after decreasing the applied magnetic
field may produce vortex motion in two directions approximately
perpendicular to each other,and both perpendicular to the applied magnetic
field. When H$||$TB, one direction of repulsion produces a motion that drives the vortices to cross the TBs, and the other direction produce a motion that drives the vortices to move between the TBs plane. Then, when H$||$TB and the field is decreasing, the
TB barriers prevent the vortices crossing the TBs, and the vortex motion
occurs preferentially between the TBs (in this sense, one might consider
that the TBs planes act as channels for exit of vortices). When H$\perp $TB,
the directions of the repulsion between vortex lines are both paralell to
the TBs planes, but in this case, a vortex line is strongly pinned by TBs
oriented  perpendicular to the vortex line. Furthermore, when field is
decreasing, one may expect that it is easier for vortices to leave the
sample for H$||$TB than for H$\perp $TB, as was observed in relaxation data
for H = 3 T. Since vortices can exit the sample easily when H$||$TB, one may
expect that below a certain applied field, the magnetization in the
decreasing-field branch for H$||$TB may eventually become smaller than the
magnetization in the decreasing-field branch for H$\perp $TB at same
temperature, as observed in MvsH curves of Fig. 2 and Fig. 3.

The values of $Hmin$, $Hon$, and $Hpen$, as obtained from the $M(H)$ curves,
are plotted in Fig. 5. The curves drawn for $Hmin$ and $Hon$ are only to
guide the eye. Below 12 K, the values of $Hmin$ increased dramatically for H$%
||$TB and became substantially larger than those for H$\perp $TB. Above 12
K, the values of $Hmin$ were approximately the same for both directions of
applied magnetic field. The values of $Hon$ were also approximately the same
for both field directions. Below 12 K, the values for H$||$TB became
measurably larger than those for H$\perp $TB. This trend was the same as
observed for $Hmin$, but the relative differences were much less.

There was little difference between the values of $Hpen$ for the two field
directions. It was found that the values of $Hpen$ for both field directions
could be fitted well by an exponential expression. An exponential response
of $Hpen$ with temperature has been observed by Andrade et al. for
anisotropic layered superconductors when H was applied along the c-axis
direction, perpendicular to the $ab$ layers \cite{Maple}. In that study, the
exponential behavior was interpreted in terms of surface barriers appearing
because of the existence of pancake-like vortices lying between the layers.
In the present study, the field was applied along the $ab$ planes and there
were no pancake-like vortices lying between the layers.

The effects of TBs on vortex phenomena can be investigated most directly
through dynamics studies. We performed flux-creep studies, emphasizing the
anisotropy of flux dynamics with respect to the TBs. Measurements
concentrated on temperatures below 12 K, because the main effects of the TBs
were observed in this temperature regime. The insets of Fig. 5 show
magnetic-relaxation curves, $M(t)$, obtained at 6 K with H = 3 T, for both
directions of applied magnetic field. The curves were obtained for the
increasing- and decreasing-field branches and for H = 0 after the field was
discharged ($Mrem$). All $M(t)$ curves presented an approximately linear
response vs. the logarithm of time. It is interesting to note the large
noise in $Mrem(t)$ and also (although not as large) in $Mout(t)$ for H$\perp 
$TB. In comparison, $Mrem(t)$ and $Mout(t)$ were quite consistent for H$||$%
TB. Such differences in noise were observed at all temperatures. The noise
in the relaxation measurements seems to be related to the resolution of the
measurement, and the fact that a vortex can exit much more easily when H$||$%
TB (large magnetic relaxation) produced less noise in this case. The insets
of Fig. 5 also reveal large differences between $Min(t)$ and $Mout(t)$ for H$%
||$TB (but not for H$\perp $TB). The insets of Fig. 5 also reveal a large
difference in $Min(t)$ and $Mout(t)$ values between H$||$TB and H$\perp $TB.

We estimated the current densities, J, at 6 K generated during the initial
stage of magnetic relaxation M(t =0) until M(t = 60-120 s) for the curves
shown in the insets of Fig. 5. For a fixed field H applied along the z
direction, in Gaussian units, $\partial B_{z}/\partial x=-4\pi J/c\sim 4\pi
\partial (M-Meq)/\partial x\sim 4\pi (dM/dt)(dt/dx)$ , where $Meq$ is the
equilibrium magnetization, and $dx/dt$ is the flux velocity (which is on the
order of $cm/s$)\cite{Tinkham}. Because we did not make local magnetization
measurements (our data were obtained over the entire volume of the sample),
we consider that $dMeq/dx$ $\sim $ $0$. From the experimental values of $%
dM/dt$ (in $emu/cm^{3}s$) and by assuming $dx/dt=1$ $cm/s$, the estimated
values of J were:

(1) H$||$TB: 0.41 A/cm$^{2}$ (flux $in$), 0.51 A/cm$^{2}$ (flux $out$), and
0.17 A/cm$^{2}$ ($remanent$);

(2) H$\perp $TB: 0.32 A/cm$^{2}$ (flux $in$), 0.32 A/cm$^{2}$ (flux $out$),
and 0.13 A/cm$^{2}$ ($remanent$).

A current value of 0.51 A/cm$^{2}$ (the largest current density that was
estimated above) corresponds to a transport current of 5 mA across the
largest area of the crystal (1 x 1 mm face), and to 0.5 mA across the
smallest area (1 x 0.1 mm face). We note that after 30 min of relaxation, $%
dM/dt$ decayed to values 30-50 times smaller than those initially
calculated. The flux velocity would also be expected to decay accordingly 
\cite{zeldov}.

Before analyzing the rate of the magnetic relaxation, it is important to
obtain the effective activation energy, $U(M)$. According to Maley et al. 
\cite{maley}, $U(M)$ be obtained from $U(M)/k_{B}=-Tln|d(M-Meq)/dt|+Tln(B%
\upsilon a/\pi d)$ , where $\upsilon $ is the attempt frequency, $a$ is the
flux hopping distance, and $d$ is the sample thickness. The equilibrium
magnetization $Meq$ is estimated as the average (M$^{+}$ + M$^{-}$)/2, where
M$^{+}$ and M$^{-}$ are respectively the magnetization in the increasing-
and decreasing-field branches of the hysteresis curve \cite{maley}. Values
of $Meq$ were found to be less than 10\% of $M$, and we therefore plotted $%
U(M)$ vs. $|M|$ instead $|M-Meq|$, for $Min(t)$ and $Mout(t)$ and for both
directions of applied magnetic field (Fig. 6). Each set of data in Fig. 6
reflects a M(t) curve and each point ($U(M)$, $M$)) in a set was obtained by
first calculating $Tln|dM/dt|$. The final value of U(M) is obtained by
adjusting a value of the constant $C=ln(B\upsilon a/\pi d)$ that produced a
smooth fit (dotted and full lines in Fig. 6) to the data obtained for a
given configuration and fixed magnetic field. For all curves of Fig. 6, $%
C=12 $. This same value of $C=12$ was reported previously for a YBaCuO
single crystal \cite{donglu}.

For $C=12$, $d$ = 0.1 mm (the thickness of the crystal), and $B\sim 3T$, $%
\upsilon a\sim 4.7cm$ $s^{-1}$, which is consistent with a flux hopping
distance $a=10$ $nm$ and an attempt frequency $\upsilon =4.7$x$10^{6}Hz$.
Each line in Fig. 6 represents a fit of $U\sim ln|M|$. With the exception of 
$U(Mout)$ for H$||$TB, data from 4 to 12 K for the other 3 configurations
fall very close to the respective $ln|M|$ line. Logarithmic decreasing of $U$
with increasing $M$ is consistent with the linear dependence of $M$ with the
logarithm of $time$, as was observed. Data at T = 2 K, the lowest
temperature of measurement, did not follow the smooth logarithmic fit, and
is not shown. Figure 6 reveals that the responses of $U(Min)$ and $U(Mout)$
were quite similar for H$\perp $TB, (in the sense that data follow the $%
ln|M| $ behavior) but not for H$||$TB for which only $U(Min)$ follow the
logarithimic behavior with M. The differences for H$||$TB explain the
differences in $Min(t)$ and $Mout(t)$ (and also in the rate of relaxation
discussed below) observed for H$||$TB, as shown in the inset of Fig. 5.

One may obtain graphically the so-called apparent pinning energy, $U_{0}$,
by constructing a tangent to a given data set in a given $U(M)$ curve (such
as in Fig. 6), where $U_{0}$ is the value at which the tangent intercepts
the $y$ axis. This value can also be obtained by the expression $U_{0}$ = -k$%
_{B}$T/$S$ , in which $S$ = (1/M$_{0}$)(d%
\mbox{$\vert$}%
M%
\mbox{$\vert$}%
/dlnt) , and $S$ is the relaxation rate and M$_{0}$ is the magnetization at
time $t=0$ \cite{beasley}. The $U_{0}$ value obtained by such a graphical
means is 50\% higher than $U_{0}$ = -k$_{B}$T/$S$, which may likely indicate
that $U_{0}$ is not well defined for the crystal used in this study.

Because of this consideration, instead of $U_{0}$, we examined the
relaxation rates S for $Min$ ($Sin$) and $Mout$ $(Sout$). The inset of Fig.
6 contains plots of $Sin$ and $Sout$ for both directions of applied magnetic
field. In this inset, the symbols $||in$ and $||out$ denote $Sin$ and $Sout$
for H$||$TB, and $\perp $in and $\perp $out denote $Sin$ and $Sout$ for H$%
\perp $TB. As in Fig. 6 proper, the curves for $Sin$ and $Sout$ for H$\perp $%
TB were similar, which provides further evidence of the flux dynamics being
approximately independent of the hysteresis branch (increasing or
decreasing) in this test configuration. On the other hand the responses vs.
temperature of $Sin$ and $Sout$ for H$||$TB confirmed that the barrier for
vortices leaving the sample was lower for this configuration. Comparison
between $Sout$ values for both configurations also indicated that the
barrier for vortices leaving the sample was lower for H$||$TB than for H$%
\perp $TB.

The relaxation rates with increasing field in the H$||$TB configuration were
smaller than in the H$\perp $TB configuration. This difference might be
related to the changes in the position of $Hmin$ observed for H$||$TB below
12 K.

The differences between various sets of curves disappeared as T approached
12 K; above 12 K, the $M(H)$ curves were quite similar for both TB
configurations. These results suggest that the TBs had a weaker effect on
flux dynamics above 12 K. It is interesting to note that surface barriers
are expected to produce a similar asymmetric, but inverted, responses for $%
Sin$ and $Sout$ as shown in the inset of Fig. 6 for the configuration H$||$%
TB \cite{Burlachkov}.

We finally discuss the flux jumps observed at low temperatures. The data at
2 K (Fig. 3) show the larger magnitude of the flux jumps when H$||$TB, and
the data at 4 K (inset of Fig. 3), for which a flux jump only occurred in
the hysteresis curve for H$||$TB, show the same trend. No flux jumps
occurred above 4 K (Fig. 2 and other data not presented in the figures).

It is likely that flux jumps appear due to a continuous imposed $dH/dt$. On
the other hand, data were collected at a fixed value of H which was reached
for a fixed value of $dH/dt$ = 0.02 $T/s$. In the field region in which flux
jumps were observed, magnetization was measured in intervals of $0.3$ $T$.
Upon analyzing the time between two consecutive data points, with $\Delta $H
= 0.3 T, we concluded that the value of $0.02$ $T/s$ was not achieved in an
interval of $0.3$ $T$. The built-in program used to charge the magnet
probably increased $dH/dt$ to a maximum value, which depended on $\Delta $H,
and then rapidly decreased $dH/dt$ as $\Delta $H approached its limit. (In
the analysis, we assumed that the ramping of $dH/dt$ was reproducible for a
given $\Delta $H.)

Magnetization was measured at a fixed field, i.e., $dH/dt=0$, and therefore
a given flux jump might appear smaller than it would if the value of $\Delta 
$H were larger. For this reason, we did not repeat the measurements with
larger values of $\Delta $H.

Obtaining data with various values of $dH/dt$ is desirable once flux jumps
(size, number, and value of the first applied field at which one occurs) are
dependent on $dH/dt$. From Fig. 3 one may observe that after a flux jump
occurred, a few increments of $\Delta $H were required for the curve to
return to an extrapolation of its previous path. This fact suggests that the
actual size of each flux jump (as would be obtained from an experiment with
a continuously varying field $dH/dt$) was close to the size that was
observed in our experiments.

The data in Fig. 3 strongly suggest that there was flux-jump anisotropy with
respect to the TBs. To our knowledge there has been no previous report of
flux jumps in twin-aligned single crystals of YBaCuO. Systematic studies of
the first flux jump and its reproducibility and size, obtained with
satisfactory statistics, are planned. This work requires continuous varying
of $dH/dt$.

We estimated the heat generated by the first flux jump at 2 K (Fig. 3). The
dissipation that heats the sample can be calculated by: $\int_{0}^{H_{1}}MdH$
= $\int_{2}^{T_{f}}C_{v}dT$ . The first expression is strictly applicable
for a total flux jump that heats a sample to the normal state, which was not
the case in our measurements. The second expression, the integral of $C_{v}$%
dT from the initial temperature to the final temperature, $T_{f}$, can be
evaluated by using published values of the specific heat of YBaCuO (T$_{c}$
= 90 K) as a function of temperature for fixed fields \cite{NIST}. By
plotting $M(H)\ $curves for a given field orientation at 2, 4, 6, and 8 K,
we obtained an approximate value for the temperature reached by the first
flux jump that occurred at 2 K. The first flux jump for H$||$TB occurred at H%
$\sim $6 T. It heated the sample to $\sim $6 K; the estimated energy of the
heating was 2.8 x 10$^{-8}$ $Joules$. The first flux jump for H$\perp $TB
occurred at H$\sim $5T. It heated the sample to $\sim $4 K; the estimated
energy of the heating was 7.3 x 10$^{-9}$ $Joules$. The heating produced
with H$||$TB was almost four times larger than that produced when H$\perp $%
TB.

Based on the these vortex-dynamics data, differences in the sizes of flux
jumps depending on wether H$||$TB or H$\perp $TB are to be expected. Because
the TBs constitute an especially strong pinning center in the configuration H%
$||$TB, a vortex can be pinned along its entire length. Under increasing
magnetic pressure, the vortices eventually pour in, similar to an avalanche,
which may induce a heating instability. In contrast, with H$\perp $TB, the
vortices pass through a given TB at one point only, forming a weaker pinning
center and inducing a smaller effect on the vortex dynamics. Indeed, from
Fig. 3 it is clear that flux jumps in the H$\perp $TB configuration were
smaller than in the H$||$TB configuration, and at 4 K the H$\perp $TB
configuration exhibited a bulk-like response.

$Conclusions$

Isothermal M$(H)$ curves exhibited two minimums below 38 K. Above 12 K, the
field position of the minimums for the H$||$TB and H$\perp $TB curves were
quite similar. Below 12 K, the position of the second minimum occurred at a
higher field value with H$||$TB. Below 6 K, only one minimum appeared for
both field directions. At low temperatures, maximums in J$_{c}$ occurred at
the positions of the minimums in the $M(H)$ curves. Vortex lines were
expelled more easily for H$||$TB than for H$\perp $TB. At T 
\mbox{$<$}%
12 K with H$||$TB, the relaxation rate for flux lines leaving the crystal
was different from that for flux entering the crystal.

We also observed flux jumps in the magnetization curves at low temperatures,
with their sizes depending wether H was applied parallel or perpendicular to
the TBs. The studies indicated that TBs may act as oriented defects
producing anisotropic flux jumping as vortex lines moved inside the sample.

Acknowledgments: We want to acknowledge J. L. Tholence and E. H. Brandt for
helpfull discussions and W. Kwok for helpfull discussions and for suggesting
the small angle dependence experiment. The work at Argonne National
Laboratory was supported by the U.S. Department of Energy, under Contract
W-31-109-Eng-38. Work supported by CNPq and FAPESP, brazilian agencies.


\begin{references}
\bibitem{stern}  M. Sarikaya and E. A. Stern, Phys. Rev. B37, 9373 (1988).

\bibitem{bakel}  G. P. E. M. van Bakel, P. A. Hof, J. P. M. van Engelen, P.
M. Bronsveld, and J. Th. M. De Hosson, Phys. Rev. B41, 9502 (1990).

\bibitem{liu}  J.Z. Liu, Y. X. Jia, R. N. Shelton, and M. J. Fluss, Phys.
Rev. Lett. 66, 1354 (1991)

\bibitem{swartzendruber}  L. J. Swartzendruber, A. Roitburd, D. L. Kaiser,
F. W. Gayle, And L. H. Bennett, Phys. Rev. Lett. 64, 483 (1990)

\bibitem{kwok}  W.K. Kwok, U. Welp, G. W. Crabtree, K. G. Vandervoort, R.
Hulscher, and J. Z. Liu, Phys. Rev. Lett. 64, 966 (1990)

\bibitem{duran}  C. A. Duran, P. L. Gammel, R. Wolfe, V. J. Fratello, D. J.
Bishop, J. P. Rice, and D. M. Ginsberg, Nature 357, 474 (1992)

\bibitem{gyorgy}  E. M. Gyorgy. R. B. van Dover, L. F. Schneemeyer, A. E.
White, H. M. O'Bryan, R. J. Felder, J. V. Waszczak, and W. W. Rhodes, Appl.
Phys. Lett. 56, 2465 (1990)

\bibitem{ousseana}  M. Oussena, P. A. J. de Groot, S. J. Porter, R. Gagnon,
and L. Taillefer, Phys. Rev. B51, 1389 (1995); M. Oussena, P. A. J. de
Groot, K. Deligiannis, A. V. Volkozub, R. Gagnon, and L. Taillefer, Phys.
Rev. Lett. 76, 2559 (1996)

\bibitem{vlasko-vlasov}  V.K. Vlasko-Vlasov, L. A. Dorosinskii, A. A.
Polyanskii, V. I. Nikitenko, U. Welp, B. W. Veal, and G. W. Crabtree, Phys.
Rev. Lett. 72, 3246 (1994)

\bibitem{rinke}  Rinke J. Wijngaarden, R. Griessen, J. Fendrich, and W. K.
Kwok, Phys. Rev. B55, 3268 (1997)

\bibitem{duran2}  C. A. Duran, P. L. Gammel, D. J. Bishop, J. P. Rice, and
D. M. Ginsberg, Phys. Rev. Lett. 74, 3712 (1995)

\bibitem{pastoriza}  H. Pastoriza, S. Candia, and G. Nieva, Phys. Rev. Lett.
83, 1026 (1999)

\bibitem{herbsommer}  J. A. Herbsommer, G. Nieva, and J. Luzuriaga, Phys.
Rev. B62, 3534 (2000))

\bibitem{jorge}  G. A. Jorge and E. Rodriguez, Phys. Rev. B61, 103 (2000)

\bibitem{bondareko}  A. V. Bondareko, et al., Low Temp. Phys. 27, 339
(2001); ibid, Low Temp. Phys. 27, 201 (2001)

\bibitem{esquinazi}  P. Esquinazi, A. Setzer, D. Fuchs, Y. Kopelevich, E.
Zeldov, and C. Assmann, Phys. Rev. B {\tt 60}, 12454 (1999)

\bibitem{mints}  R. G. Mints and E. H. Brandt, Phys. Rev. B 54, 12421 (1996)

\bibitem{muller}  K.-H. Muller and C. Andrikidis, Phys. Rev. B 49, 1294
(1994)

\bibitem{guillot}  M. Guillot, M. Potel, P. Gougeon, H. Noel, J. C. Levet,
G. Chouteau, and J. L. Tholence, Physics Letters A 127, 363 (1988)

\bibitem{reply}  S. Salem-Sugui Jr., A. D. Alvarenga, M. Friesen, B. Veal,
and P. Paulikas, Phys. Rev. B 63, 216502, (2001)

\bibitem{beam}  C. P. Bean, Phys. Rev. Lett. 8, 250 (1962)

\bibitem{Tinkham}  Michael Tinkham, Introduction to Superconductivity,
second edition, McGraw-Hill, Inc. 1996

\bibitem{Maple}  M.C. de Andrade, N. R. Dilley, F. Ruess, and M. B. Maple,
Phys. Rev. B57, R708 (1998)

\bibitem{zeldov}  Y. Abulafia, A. Shaulov, Y. Wolfus, R. Prozorov, L.
Burlachkov, Y. Yeshurun, D. Majer, E. Zeldov, and V. M. Vinokur, Phys. Rev.
Lett. 75, 2404 (1995)

\bibitem{maley}  M. P. Maley, J. O. Willis, H. Lessure, and M. E. McHenry,
Phys. Rev. B42, R2639 (1990)

\bibitem{donglu}  D. Shi and S. Salem-Sugui, Jr., Phys. Rev. B 44, 7647
(1991)

\bibitem{beasley}  M. R. Beasley, R. Labash and W. W. Weeb, Phys. Rev. 181,
682 (1969)

\bibitem{Burlachkov}  L. Burlachkov, Phys. Rev. B47, 8056 (1993); A.D.
Alvarenga and S. Salem-Sugui, Jr., Physica C 235, 2811 (1994)

\bibitem{NIST}  NIST WebHTS Data base (A. Junod et al., Physica C 162-164
(1989); G. Triscone et al., Physica C 168, 40 (1990); J. Y. Genoud et al.,
Physica C 177, 315 (1991))

Fig. 1. Microscopic photo of the sample (100x enlarged), evidencing paralell
twin boundaries along the surface. The photo was obtained by adjusting the incident angle of the light into the sample's surface to the Brewster angle allowing the observation of the twin planes orientation.

Fig. 2. Selected M-vs.-H curves at 25, 15, and 8 K for H$||$TB and H$\perp $%
TB; insets show curves at 8 K after changing the angle between H and the TBs
by $\sim 7{{}^{\circ }}$.

Fig. 3. M-vs.-H curves for H$||$TB and H$\perp $TB at 2 K (inset at 4 K).

Fig. 4 $J_{c}vsH$ curves, as estimated from $M(H)$ curves obtained below 10
K for $H||TB$ and $H\perp TB$.

Fig. 5. $Hpen$, $Hon$, and H$min$ vs. $T$; insets show magnetic-relaxation
curves at 6 K and H = 3 T ($in$ and $out$) and for H = 0 ($rem$) for H$||$TB
and H$\perp $TB.

Fig. 6 Effective activation energy, $U$, for flux $in$ and flux $out$ for H
= 3 T vs. 
\mbox{$\vert$}%
$M|$ for H$||$TB and H$\perp $TB. Solid and dotted lines represent fits of $%
U\sim ln|M|$ and inset shows rate of relaxation $S$ $vs.$ $T$ for
relaxation-data set obtained for H = 3 T.
\end{references}
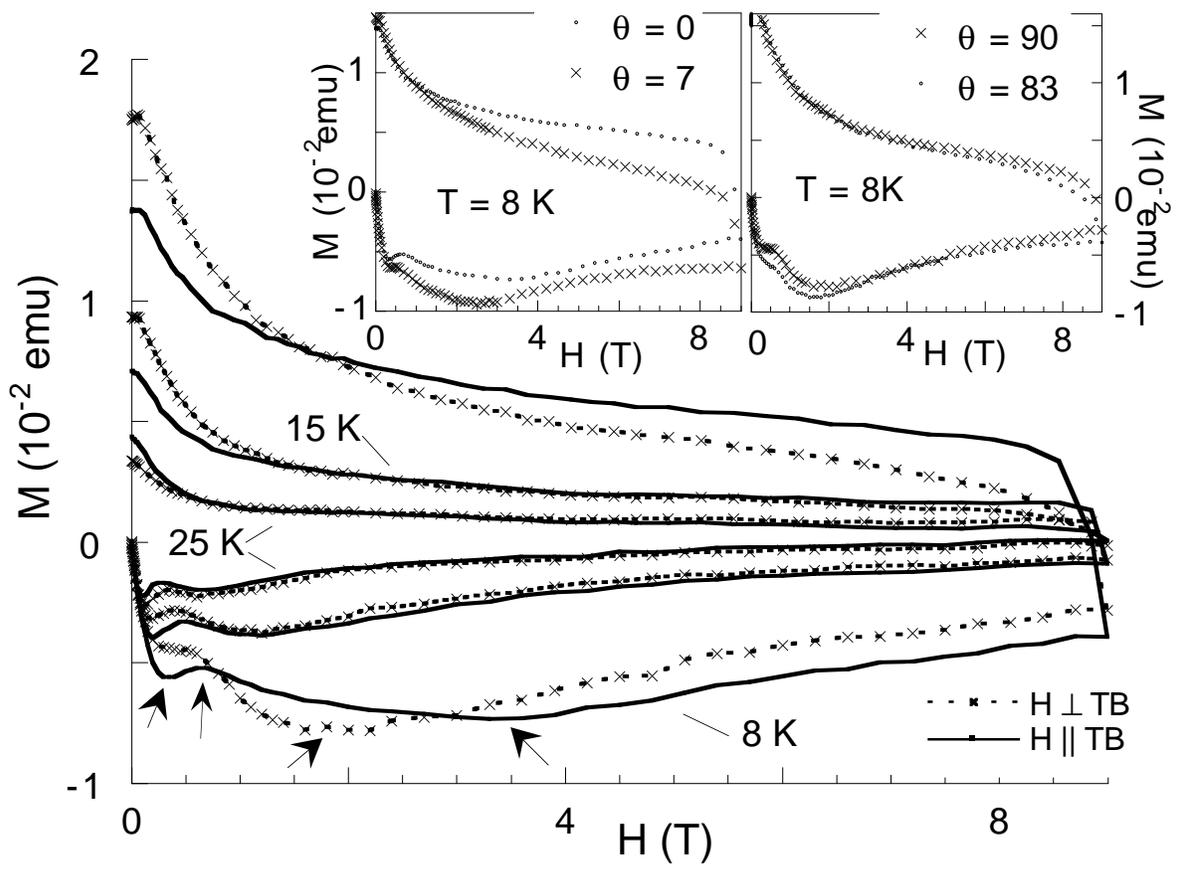
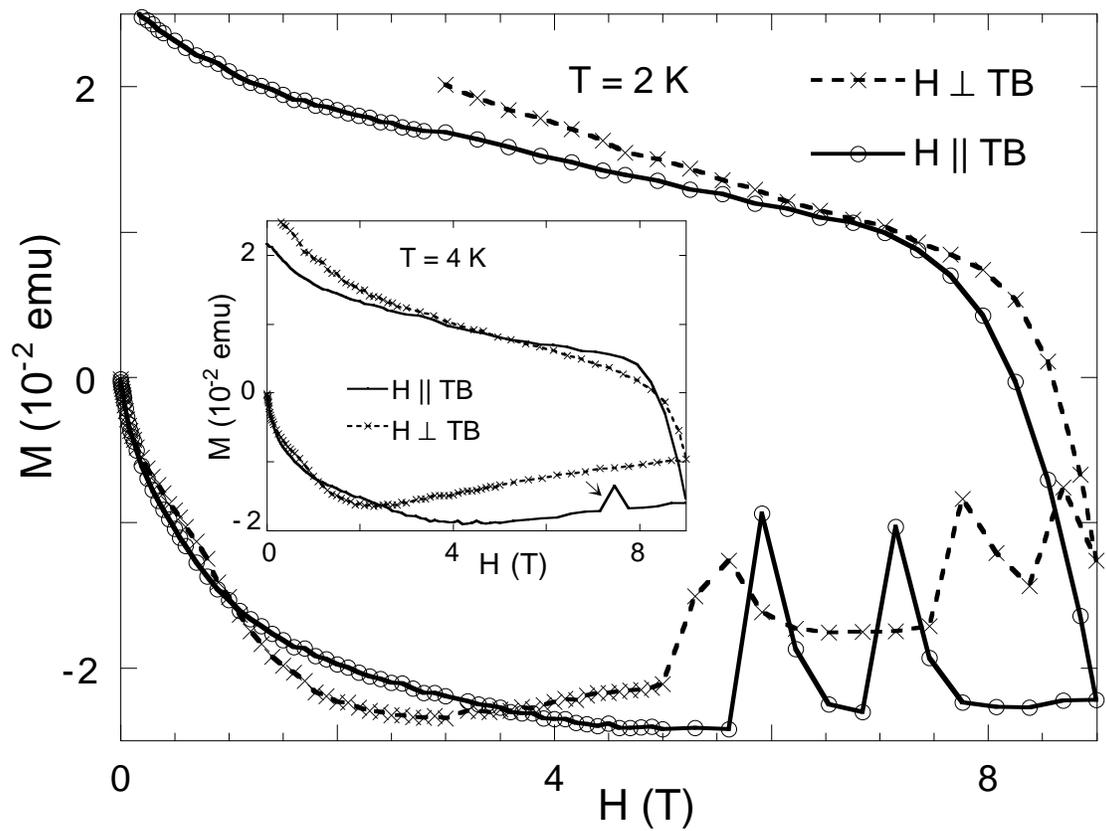
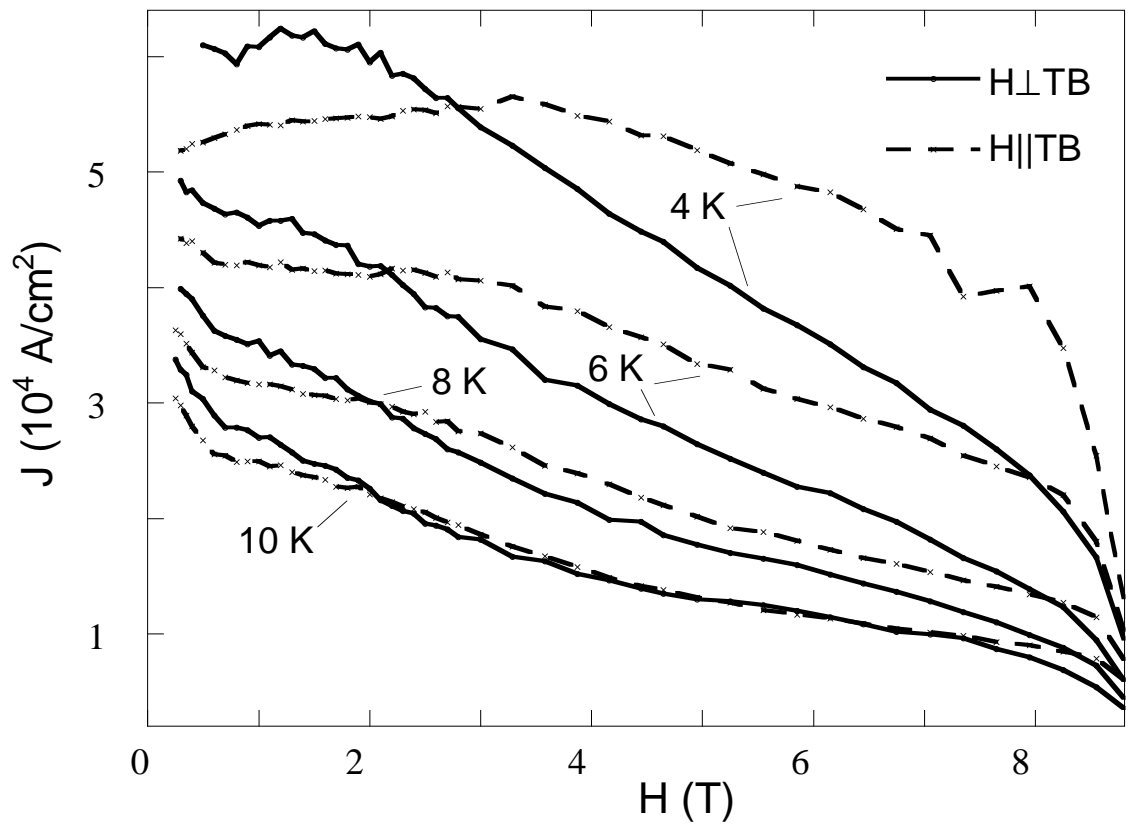
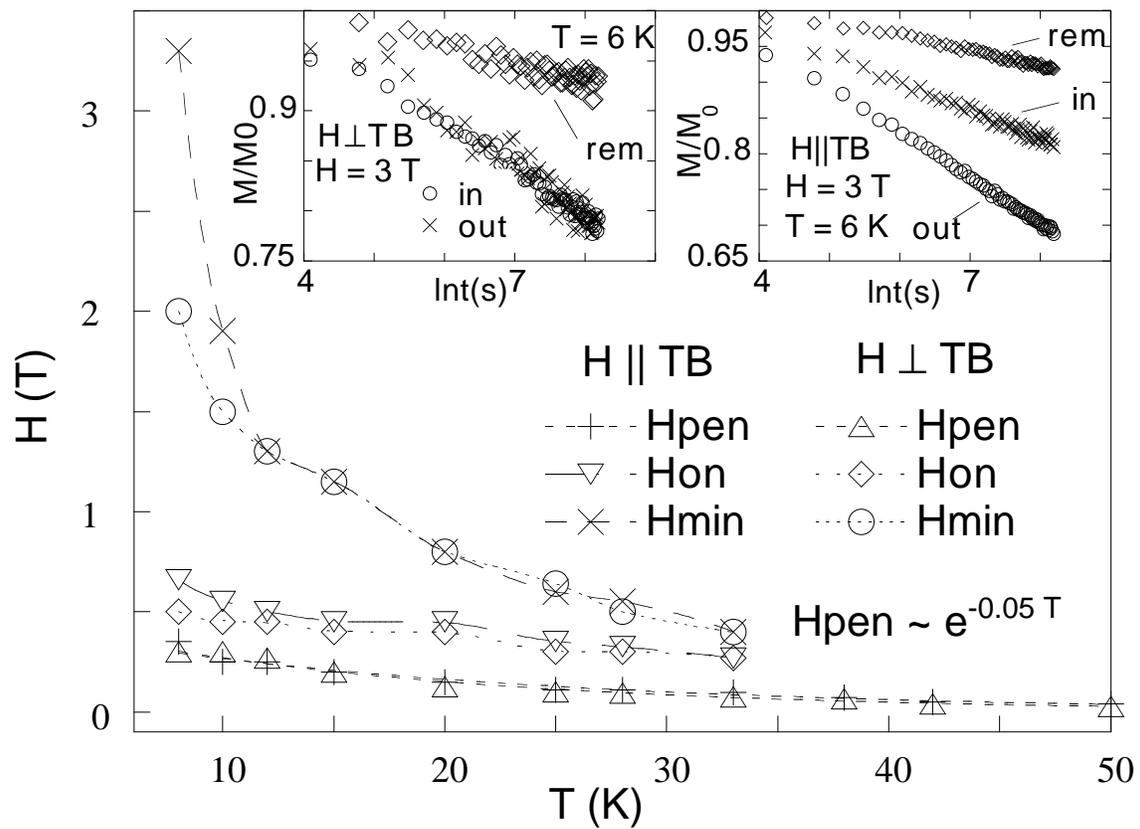
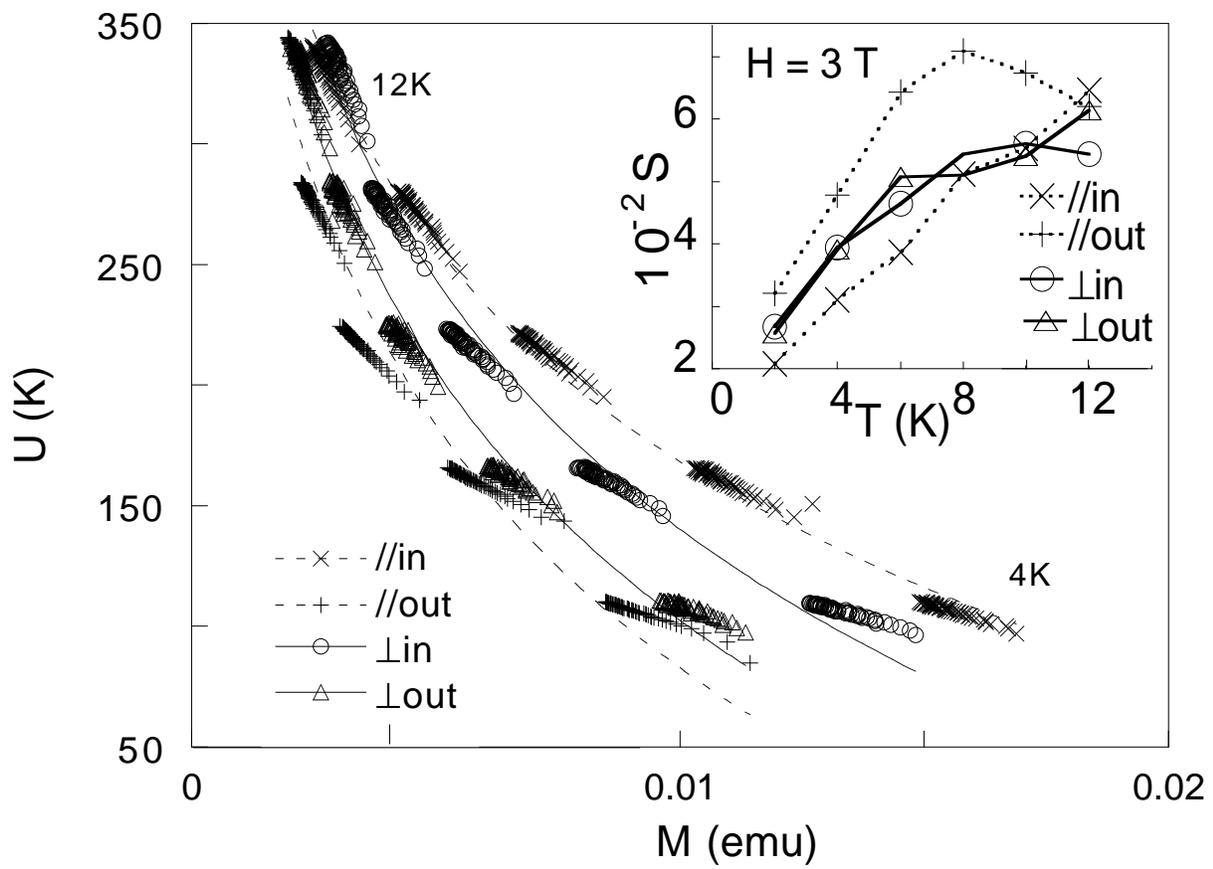
\end{document}